\documentclass[aps,pre,twocolumn,superscriptaddress]{revtex4-1}

\usepackage{graphicx}
\graphicspath{ {./Figures/} }
\usepackage{amsmath}
\usepackage{upgreek}
\usepackage{textcomp,amssymb}
\usepackage{hyperref}

\usepackage{tikz}

\usepackage{xcolor}

\newcommand{\Rout}{R_\text{out}}
\newcommand{\Rin}{R_\text{in}}
\newcommand{\Vcav}{V_{\text{cav}}}
\newcommand{\Vrod}{V_{\text{rod}}}

\newcommand{\bvec}[1]{{\mathbf{\string#1} }}

\usepackage{mathtools}

\newcommand{\Qvec}{\mathbf{Q}}
\newcommand{\Ivec}{\mathbf{I}}
\newcommand{\Sop}{S}
\newcommand{\ofr}{(\mathbf{r})}

\usepackage[normalem]{ulem}
\begin{document}

\title{Colloidal smectics in button-like confinements: experiment and theory}

\author{Ren\'{e} Wittmann}
\email[]{rene.wittmann@hhu.de}
\affiliation{Institut f\"ur Theoretische Physik II: Weiche Materie, Heinrich-Heine-Universit\"at D\"usseldorf, Germany}

\author{Paul A.\ Monderkamp}
\affiliation{Institut f\"ur Theoretische Physik II: Weiche Materie, Heinrich-Heine-Universit\"at D\"usseldorf, Germany}

\author{Jingmin Xia}
\email[]{jingmin.xia@nudt.edu.cn}
\affiliation{College of Meteorology and Oceanography, National University of Defense Technology, China}

\author{Louis B.\ G.\ Cortes}
\affiliation{School of Applied and Engineering Physics, Cornell University,~Ithaca,~NY,~USA}
\affiliation{Physical and Theoretical Chemistry Laboratory, University of Oxford, United Kingdom}

\author{Iago Grobas}
\affiliation{Physical and Theoretical Chemistry Laboratory, University of Oxford, United Kingdom}

\author{Patrick E.\ Farrell}
\affiliation{Mathematical Institute, University of Oxford, United Kingdom}

\author{Dirk G.\ A.\ L.\ Aarts}
\email[]{dirk.aarts@chem.ox.ac.uk}
\affiliation{Physical and Theoretical Chemistry Laboratory, University of Oxford, United Kingdom}

\author{Hartmut L\"owen}
\email[]{Hartmut.Loewen@uni-duesseldorf.de}
\affiliation{Institut f\"ur Theoretische Physik II: Weiche Materie, Heinrich-Heine-Universit\"at D\"usseldorf, Germany}

\date{\today}

\begin{abstract}
Liquid crystals can self-organize into a layered smectic phase.
While the smectic layers are typically straight forming a lamellar pattern in bulk, external confinement may drastically distort the layers due to the boundary conditions imposed on the orientational director field.
Resolving this distortion leads to complex structures with topological defects.
Here, we explore the configurations adopted by two-dimensional colloidal smectics made from nearly hard rod-like particles in complex confinements, characterized by a button-like structure with two internal boundaries (inclusions): a two-holed disk and a double annulus.
The topology of the confinement generates new structures which we classify in reference to previous work as generalized laminar and generalized Shubnikov states.
To explore these configurations, we combine particle-resolved experiments on colloidal rods with three complementary theoretical approaches:
Monte-Carlo simulation, first-principles density functional theory and phenomenological $\Qvec$-tensor modeling.
This yields a consistent and comprehensive description of the structural details.
In particular,
we characterize a nontrivial tilt angle between the direction of the layers and symmetry axes of the confinement.
\end{abstract}

\maketitle

\section{Introduction}

Liquid crystals \cite{PhLiqCrys,lavrentovich1998topological} have proven to be an important tool in the investigation of various topological phenomena. In nematic liquid crystals, topological defects reflect the frustrated orientational order due to, e.g., confining geometries
\cite{dammone2012,VirNemConfGeom,garlea2016finite,tran2016lassoing,han2020reduced,yaochen2020,yao2021defect,QuantSelAss,ienaga2023geometric}, active dynamics \cite{actNem,keber2014topology, decamp2015orientational, giomi2015geometry,tan2019topological} or a combination of both \cite{shendruk2017dancing,hardouin2022active,huang2022activesmectics}.
 This diversity of defects and formation pathways has led to extensive research attention devoted to understanding and accurately modeling the emerging topological defect structures \cite{toptherdef,hydroOfDef,mosna2012,nem_defects,real_defch,vromans2016orientational,tang2017orientation}.

In recent years, there has been an increasing interest in layered liquid crystals \cite{guin-2018-article} and, in particular, smectic phases \cite{liarte2015, radzihovsky2020_gaugetheorySM, paget-2022-article, zappone-2022-review}, which possess both orientational order and a periodic modulation of the center-of-mass density in the form of layers. This development owes to progress in (i) advances in experiments \cite{10.1021/acs.nanolett.9b04347,10.1073/pnas.2000849117,10.1021/acs.langmuir.5b02508,10.1002/adom.201500153,originalPointsToLines,10.1039/d0sm01112f,10.1021/acs.langmuir.7b03351,10.1007/s00396-010-2367-7,ma13173761,Gim2017,kuijk2012phase,CollLQinSqConf,granular0,armas2020},
(ii) continuum modeling \cite{pevnyi2014, ballbed-2015-article, xia-2021-article, paget2022tensorsmectic} and
(iii) first-principles theory \cite{wittmann2014,wittmann2016,PD_Rene}, complemented by
(iv) topological insight \cite{chen2009symmetry,kamien2016topology,aharoni2017composite,machon2019}  reinforced by
(v) simulating particle-resolved defect structures \cite{PhysRevE.68.021706,monderkamp2021topology,monderkamp2022topological,monderkamp2023network}.
Despite these advances,
 further interdisciplinary efforts are needed to achieve a comprehensive understanding of smectics
and to bridge the gap between different (model) systems showing smectic layering  and   theoretical approaches of all kinds.
For example, different classes of experimental smectics range from thermotropic molecular liquid crystals with highly elastic layers to colloidal smectics, whose internal structure, being governed by packing effects, is  more rigid.
A particular challenge is thus to identify and develop universal structural and topological classification concepts that are applicable to all kinds of smectic layering.

\begin{figure}[t]
\begin{center}
\includegraphics[width=0.9\linewidth]{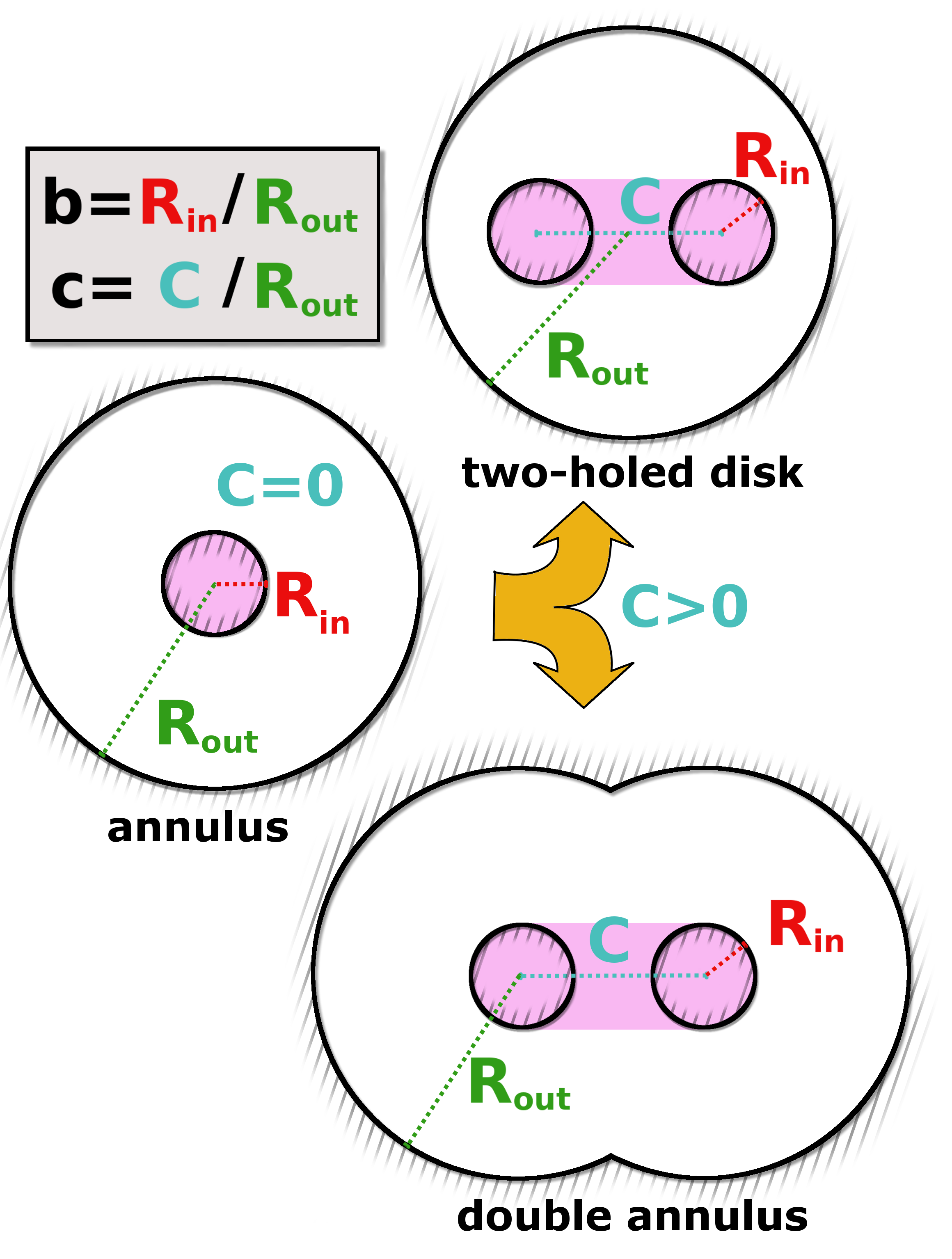}
\caption{Schematic depiction of the button-like confining geometries investigated in this work generalizing an annulus (middle) with inclusion size ratio $b=\Rin/\Rout$, where $\Rout$ is the radius of the outer confining wall and $\Rin$ is the radius of the circular wall in the interior, called inclusion \cite{annulus}.
Here, we consider two inclusions at distance $C$, introducing the inclusion distance ratio $c=C/\Rout$.
In detail, we define the two-holed-disk geometry (top), where the outer wall is always circular, and the double-annulus geometry (bottom), where the outer wall is defined via two intersecting circles at the same distance $C$.
The outer radius $\Rout$ is kept at a fixed value throughout the manuscript, which corresponds to about five smectic layers.
Further details are provided in Sec.~\ref{sec_geometry}.
}
\label{fig_concept}
\end{center}
\end{figure}

In this paper, we bring four complementary approaches together to understand the structure of colloidal smectics, confined to two-dimensional domains with a complex topology involving two holes.
We use particle-resolved colloidal experiments on silica rods \cite{CollLQinSqConf}, Monte-Carlo simulations of a hard-rod model \cite{monderkamp2021topology}, microscopic density functional theory (DFT) for hard rods \cite{PD_Rene} and a recent continuum $\Qvec$-tensor model extending the Landau--de Gennes theory for nematics \cite{xia-2021-article}.

 The goal of our work is twofold.
First, we  provide a more profound topological understanding of colloidal smectics by exploring more complex confinements that allow for a larger structural variety.
Second, we bring together four different approaches to both demonstrate the general applicability of our topological concepts and exploit their synergies when it comes to a detailed structural analysis.
We will thus demonstrate that our methods not only yield consistent predictions of the properties of different topological states but also allow us to tackle the problems from different viewpoints.
In particular, we investigate responses of the system to both changes of the confining geometry and variation of interaction parameters.
This allows us to systematically explore the ranges of stability of different topological states as well as different alignment phenomena and structures in the absence of a continuous rotational symmetry.

 This manuscript is arranged as follows. We provide details on the confining geometries, methods and topological classification scheme in Sec.~\ref{sec_allmethods}, before discussing our results in Sec.~\ref{sec_results}.
After summarizing our observations, we conclude in Sec.~\ref{sec_conclusions}.

\section{Confined smectic states \label{sec_allmethods}}

Our goal is to identify the structure and topology of the  smectic states emerging in various two-dimensional confinements.
The complexity of a confining domain can be both of explicit geometrical origin, related to the curvature of the walls, or of topological origin.
The latter can be quantified by the Euler characteristic $\chi$, which counts the number of connected components minus the number of holes in two dimensions, irrespective of the particular geometric shape of the domain.
In the following, we elaborate on the relevant geometrical parameters (Sec.~\ref{sec_geometry}), describe how we resolve smectic structures in experiments and three theoretical approaches (Sec.~\ref{sec_methods}) and provide details on the topological analysis (Sec.~\ref{sec_classification}).

\subsection{Button-like confinements \label{sec_geometry}}

As confining domains, we consider two generalizations of an annular geometry (central drawing in Fig.~\ref{fig_concept}). This allows us to compare against solutions characterized for an annulus in previous work \cite{annulus}.
An annulus is composed of a large circle of radius $\Rout > 0$ with a single circular hole (inclusion) of radius $\Rin < \Rout$ in the center.
This topology has an Euler characteristic $\chi=0$,
which allows for a structure free of orientational topological defects at the cost of forming an array of edge dislocations to relax the deformation of the layers imposed by the curved confinement (Shubnikov state, named following the structural analogy to type-II superconductors~\cite{10.1073/pnas.2000849117,de1972analogy}).
Depending on the particular geometry, this structure competes, among others, with an undeformed structure (laminar state), which comes at the cost of the formation of grain boundaries, i.e., defects in both orientational and positional order \cite{annulus}.

The central geometrical parameter which determines the stability of a structure emerging in annular geometry is the inclusion size ratio $b=\Rin/\Rout$.
For each of our two related geometries, further specified below, we add a second inclusion of the same inclusion size ratio $b$ and
introduce the geometrical parameter $C=c\Rout$,
which denotes the distance between the centers of the two inclusions.
In general, for small enough relative distance $c\leq2b$ the inclusions intersect, resulting in a distorted annulus with an effectively stretched inclusion ($\chi=0$),
while for larger distances $c$, there are two separated holes such that the Euler characteristic equals $\chi=-1$.

First, we consider the two-holed-disk geometry (top drawing in Fig.~\ref{fig_concept}), made of a single outer circle of radius $\Rout$, and two inclusions whose centers are shifted away from each other in opposite directions with the mutual distance $C$,
such that the distance to the center of the outer disk is $C/2$ in each case.
Regarding the region accessible to the particles, the shortest distance between outer and inner walls varies upon circling along the outer wall.
Note that, for extreme distances $c\geq 2-2b$, the inclusions are in contact with the outer wall, such that the geometry becomes simply connected again ($\chi=1$).

Second, we consider the double-annulus geometry (bottom drawing in Fig.~\ref{fig_concept}), composed of two outer circles,
which are shifted alongside the two inner circles, such that each pair of inner and outer circles has the same center
with a mutual distance of $C$.
Consequently, each point of the outer confining wall has the same shortest distance $(1-b)\Rout$ to one of the inclusions.

To summarize, each geometry in Fig.~\ref{fig_concept} is fully determined by three parameters: the inclusion size ratio $b$, the inclusion distance ratio $c$ and the total size of the confining domain specified by the radius $\Rout$ of the outer wall.
Throughout the manuscript, we keep $\Rout\approx5\lambda$ fixed, where $\lambda$ is the layer spacing of the smectic.

\begin{figure}[t]
\centering
\includegraphics[width=\linewidth]{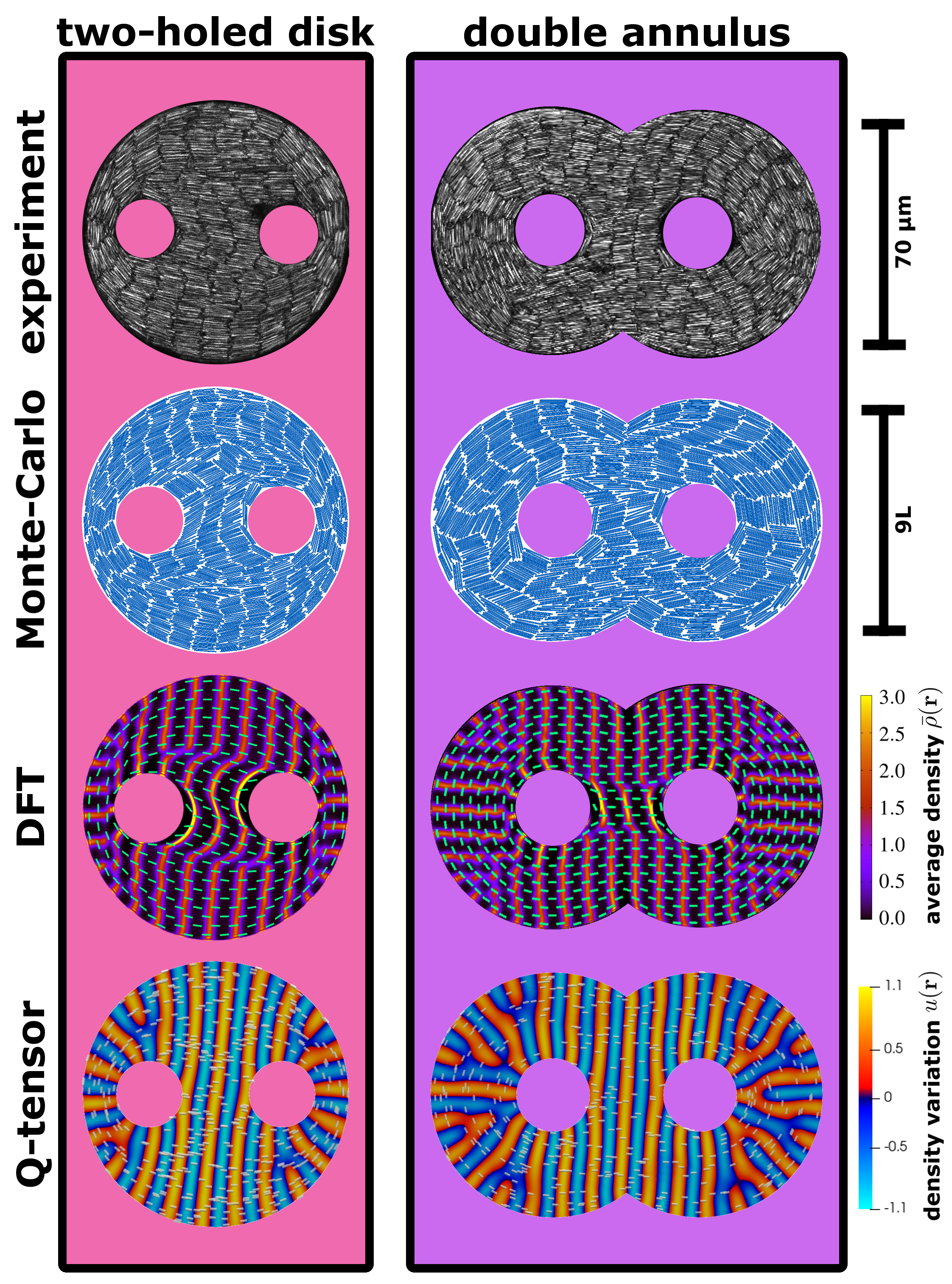}
\caption{Example structures in the two confining geometries shown in Fig.~\ref{fig_concept} with two inclusions.
The left and right columns depict the two-holed-disk geometry for $b\approx0.25$ and $c\approx1.0$
and the double-annulus geometry for $b\approx0.3$ and $c\approx1.2$, respectively.
In both geometries we show (from top to bottom) typical particle configurations in colloidal experiments, Monte-Carlo simulations of rod-like particles, density profiles and director fields from hard-rod density functional theory (DFT) and $\Qvec$-tensor theory with suitably adapted parameters (see text).}
\label{fig_overview}
\end{figure}

\subsection{Complementary
methods \label{sec_methods}}

Next we briefly introduce our methods to generate two-dimensional smectic states in extreme confinement and describe how to interpret the final smectic structures depicted in  Fig.~\ref{fig_overview}.
In general, smectic order is characterized by positional order of the particle centers in equidistant layers and orientational order along a director at a constant angle to these layers.
Confinement induces frustration of this preferred alignment in the form of deformations or discontinuities of the layers and/or the director field.
The complementary use of our different approaches, allows us to optimally exploit their advantages when it comes to understanding the driving forces behind the formation, the stability range and the topology of the emerging structures.
Further details on each method are provided in Appendix~\ref{app_methods}.

Our experiments exploit the sedimentation equilibrium of silica rods in an aqueous solution, as described in Appendix~\ref{app_methods_exp}.
The particle-resolved optical micrographs, displayed in the first row of Fig.~\ref{fig_overview}, are then taken from the bottom, where the rods settle within tailor-made cavities.
These quasi-two-dimensional smectic structures can then be analyzed by direct optical inspection or
reading out individual particle coordinates and orientations from processed images.

We also perform Monte-Carlo simulations on confined systems of rods modeled as hard discorectangles in the canonical ensemble, as described in Appendix~\ref{app_methods_sim}.
The simulation snapshots, as displayed in the second row of Fig.~\ref{fig_overview}, can be analyzed in the same manner as those from the experiments, while
this particle-resolved numerical method leaves us in full control of the particle shape, number density and geometrical parameters.
 This allows us to gather a large amount of statistics for any prescribed geometry.
From that we can further locate the grain boundaries by sampling a local version (cf.\ Fig.~\ref{fig_FAinclusions}) of the two-dimensional orientational order parameter $\Sop\ofr =  | \left < \exp \left( i2\phi\ofr\right )\right > |$, where $\phi$ denotes the orientation angle of the individual rods within a local environment around the position $\mathbf{r}$.

On the theory side, we employ classical density functional theory (DFT) \cite{Evans1979} for hard discorectangles, as described in Appendix~\ref{app_methods_DFT}.
In DFT, all structural information is comprised within the number density $\rho(\mathbf{r},\phi)$ found by minimizing an appropriate functional $\Omega[\rho]$.
This central quantity reflects the probability of finding a particle with the center-of-mass position $\mathbf{r}$ and its long axis oriented in a direction given by the angle $\phi$.
The typical density profiles, as displayed in the third row of Fig.~\ref{fig_overview}, indicate both the smectic layers by a color plot of the local density (averaged over all orientations)
and the director by green bars.
In the employed version of DFT \cite{PD_Rene,annulus} based on fundamental measure theory \cite{Rosenfeld1989,roth2010fundamental,Rosenfeld1989}
 the interactions are treated on a microscopic level through the geometry of individual particles, such that the density profiles reflect the particle dimensions.
 As DFT is founded in statistical mechanics, no additional averaging is required and
the most stable state can be identified among multiple solutions from the minimal value of the corresponding free energy.

Furthermore, we study a recent phenomenological model for smectic layering, based on an extension of Landau--de Gennes theory to smectics, as described in Appendix~\ref{app_methods_LdG}.
It minimizes a total free energy $\mathcal{J}(u, \Qvec)$ of the local density perturbation $u(\mathbf{r})$ for smectic phases and a tensorial order parameter $\Qvec(\mathbf{r})$ encoding the orientational order.
As displayed in the fourth row of Fig.~\ref{fig_overview}, typical profiles of the smectic density variation $u$ exhibit maxima (light yellow) and minima (light blue) which both can be interpreted as the smectic layers, while the orientational director field (gray rods) corresponds to the eigenvector of tensor $\Qvec$ with the largest eigenvalue.
In this smectic $\Qvec$-tensor theory, the interactions are implicitly described through  a range of phenomenological parameters.
To connect to our other approaches, we consider two of these parameters as free variables:
the elasticity parameter $K$ of the director field and the anchoring parameter $w$, which indicates the strength of tangential alignment of directors at the outer wall.
As the latter depends on curvature, the alignment at the inner wall is accordingly weaker (see Appendix~\ref{app_methods_LdG} for further details).
When choosing $K=0.5$ and $w=5$ (two-holed disk) or $K=1$ and $w=10$ (double annulus)
we find convincing agreement with our other methods in Fig.~\ref{fig_overview}.

\begin{figure}[t]
\centering
\includegraphics[width=1.0\linewidth]{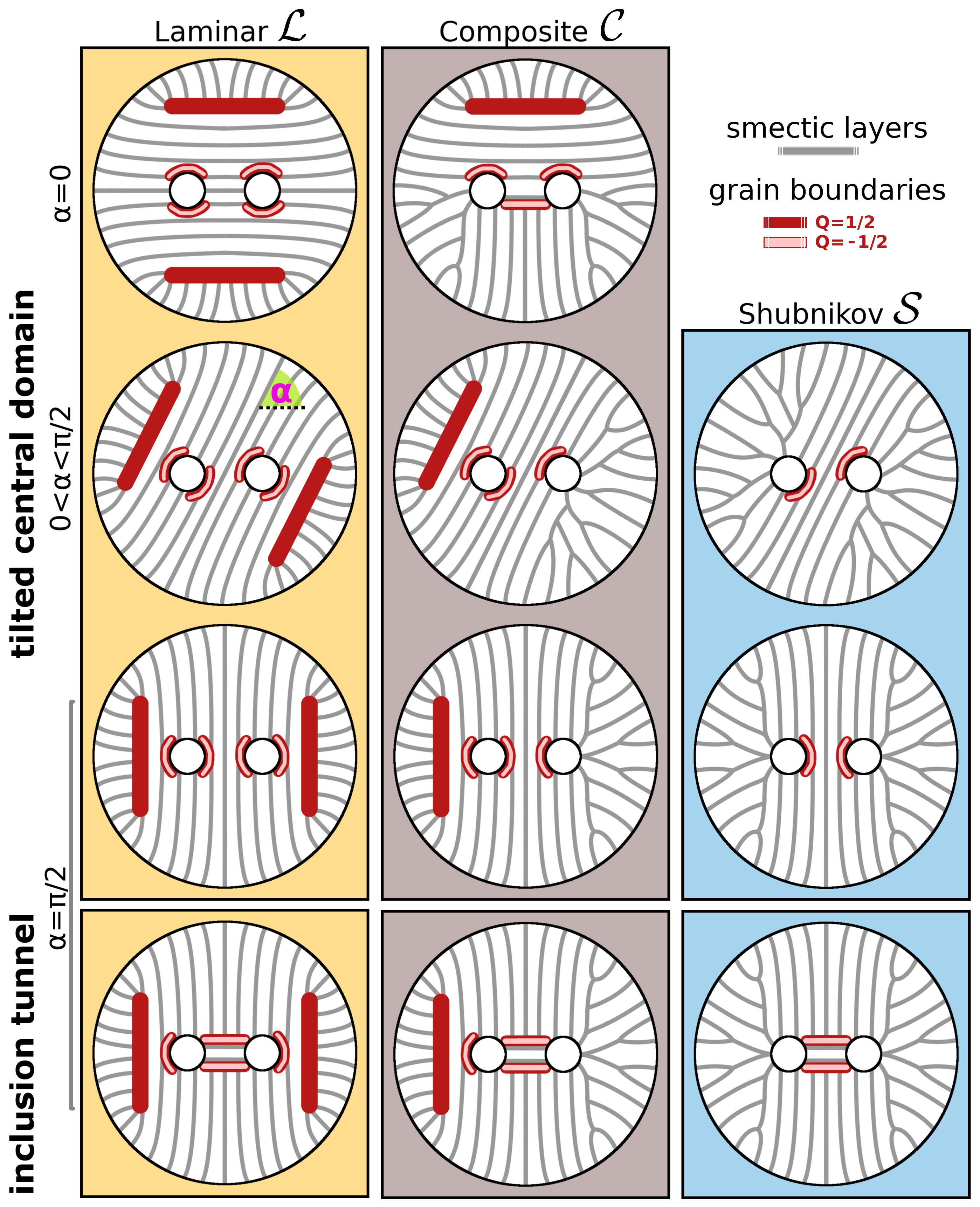}
\caption{Illustrative overview of topological structures.
The rows show different realizations of generalized laminar ($\mathcal{L}$, two $+1/2$ and four $-1/2$ charges), composite ($\mathcal{C}$, one $+1/2$ and three $-1/2$ charges) and generalized Shubnikov ($\mathcal{S}$, two $-1/2$ charges) states.
The first three rows display the structures with a continuous central domain which is tilted by the angle $\alpha$ relative to the axis that connects the two inclusion centers, as annotated (the definition of $\alpha$ is included in the first illustration of the second row). The fourth row displays the structures for $\alpha=\pi/2$ with the central domain interrupted by a few layers spanning between the two inclusions, which we refer to as inclusion tunnel.
 The illustrations depict the typical appearance of these generalized states in the two-holed-disk geometry (where we also speak of dual laminar and stretched Shubnikov states),
 but the same general classification also holds for the double-annulus geometry.
 Further details are provided in Sec.~\ref{sec_classification}.
}
\label{fig_structures}
\end{figure}

\subsection{Topological classification \label{sec_classification}}

The hard rods described in our particle-based approaches favorably align parallel to the system walls.
This externally imposed boundary condition competes with the intrinsic smectic structure favoring defect-free, undeformed, parallel and equidistant layers.
The resulting (stable or metastable) equilibrium structures are thus governed by a balance between elastic deformations and topological defects.
The type, location and shape of the emerging topological defects provide a convenient way to classify and distinguish between the observed smectic states, as illustrated in Fig.~\ref{fig_structures}.  \cite{annulus,monderkamp2021topology}.
In smectics, we typically observe spatially extended grain boundaries
or virtual boundary defects (misalignment of rods at the wall), whose orientational frustration can be quantified by a topological charge $Q$ in analogy to nematic disclinations~\cite{monderkamp2021topology,monderkamp2022topological}.
The Poincar\'e--Hopf theorem gives rise to a fundamental law of charge conservation for two-dimensional smectic structures:
the total sum of topological charges $\sum Q=\chi$ in a confined system must equal the confining domain's Euler characteristic $\chi$.
The two main types of grain boundaries relevant in our study possess a $Q=+1/2$ or a $Q=-1/2$ topological charge,
associated with a clockwise and counterclockwise rotation of the director field around the defect, respectively.
In both cases, the main rotation occurs at the end points of the grain boundaries, which can then be identified as point-like tetratic defects of quarter-integer magnitude~\cite{monderkamp2021topology}.

As a first step of our topological analysis, we focus on the structural properties on the largest scale and ignore the details in the region between the two inclusions.
By doing so, we can classify the overall smectic states in the same spirit as in an annular geometry \cite{annulus},
i.e., by considering an effective geometry with Euler characteristic $\chi_\text{eff}=0$, obtained by formally replacing the two inclusions with a single inclusion given by their convex hull (indicated by the magenta shaded areas in Fig.~\ref{fig_concept}).
 Following the nomenclature of Ref.~\cite{annulus}, we classify solutions into (i) generalized laminar states, with two $Q=+1/2$ defects close to the outer wall and two $Q=-1/2$ defects at the effective inclusion, (ii) generalized composite states, combining features of both (one $Q=+1/2$, one $Q=-1/2$ defect and edge dislocations),  or (iii) generalized Shubnikov states, with no topological charges but edge dislocations of the layers, as sketched in the different columns of Fig.~\ref{fig_structures}.
Due to the particular appearance of these states in the two-holed-disk geometry, we also speak of dual laminar and stretched Shubnikov states in this case.

As a second step, we account for the broken
continuous rotational symmetry for a nonzero inclusion distance $c>0$. We introduce the tilt angle $\alpha\in[0,\pi/2]$ of the central domain with respect to the axis that connects the two inclusion centers (cf.\  the second row in Fig.~\ref{fig_structures}) as an additional structural quantifier.
The rows of Fig.~\ref{fig_structures} depict the different states for
$\alpha=0$ (layers parallel to the connecting axis), an intermediate value of $\alpha$,
and $\alpha=\pi/2$ (layers perpendicular to the connecting axis).

The third step of our topological analysis concerns the fine structure between the inclusions, i.e., the location and shape of the two additional $Q=-1/2$ defects required by the charge conservation to match the overall Euler characteristic $\chi=-1$ in the presence of two holes.
In general, there are two possibilities.
First, the layers between the inclusions can
align with the adjacent layers outside to become part of a larger domain,
comparing the first three rows in Fig.~\ref{fig_structures}.
In this case, the two $Q=-1/2$ defects are directly located at the inclusions.
Second, if $\alpha\gtrsim\pi/4$, it is also possible that the two inclusions are connected by one or more isolated smectic layers, such that the rods in the central region fulfill the parallel wall anchoring condition.
However, this inclusion tunnel then interrupts the central domain, compare the last row in Fig.~\ref{fig_structures}, which results in two grain boundaries with $Q=-1/2$, parallel to the line connecting the inclusions.
Note that for $\alpha\lesssim\pi/4$, the anchoring condition can be fulfilled without forming an inclusion tunnel, i.e., when the layers connecting the inclusions are
 part of the defect-free central domain.

We comment that the stretched Shubnikov state in the two-holed-disk geometry cannot be realized for small tilt angles.
When decreasing the tilt angle $\alpha$ of a structure below $\alpha\lesssim\pi/4$, the central domain eventually fulfills the first criterion for an inclusion tunnel (layers connecting the inclusions), comparing with the generalized laminar state in the top-right illustration of Fig.~\ref{fig_structures} for $\alpha=0$.
In turn, a generalized Shubnikov state is only possible if the second criterion for an inclusion tunnel is also fulfilled, i.e., that there exists a larger domain outside the convex hull of the inclusions which has a tilt angle $\alpha\gtrsim\pi/4$ and is separated from the inclusion tunnel by two grain boundaries, comparing with the bottom-right illustration of Fig.~\ref{fig_structures} for $\alpha=\pi/2$.
Combining these two structures, the composite state depicted in Fig.~\ref{fig_structures} for $\alpha=0$ constitutes a particular example with two domains of comparable size.

\section{Results \label{sec_results}}

\subsection{Two-holed disk \label{sec_BU}}
We focus first on the two-holed-disk geometry. As in the annulus, the circular shape of the outer confining wall remains invariant for all choices of the geometrical parameters $b$ and $c$, where an annular geometry is recovered for $c=0$.
To quantify the emerging structures in full detail, we proceed stepwise.
First, in Sec.~\ref{sec_BU_LS}, we study the transition between the generalized laminar and Shubnikov states,
focusing only on the smectic structures outside the convex hull of the two inclusions.
Then, we use these insights to explain our general observations in experiments and Monte-Carlo simulations in Sec.~\ref{sec_BU_OBS} before quantifying the two structural aspects indicated in Fig.~\ref{fig_structures}.
In particular, we study the tilt angle $\alpha$ of the central smectic domain in Sec.~\ref{sec_BU_OR} and investigate the locations of the topological defects
in the region between the inclusions in Sec.~\ref{sec_BU_FS}.

\subsubsection{Theoretical laminar-Shubnikov transition \label{sec_BU_LS}}

We start by systematically mapping out a simple state diagram using DFT, which gives us full control over the structures we wish to compare.
In particular, to understand the general structural response upon varying both $b$ and $c$, we focus on (dual) laminar and (stretched) Shubnikov states, as specified in Sec.~\ref{sec_classification}.
Moreover, we restrict ourselves to (fairly) axially symmetric structures, i.e., we impose the two extreme tilt angles $\alpha=0$ or $\alpha=\pi/2$ in the laminar case and just $\alpha=\pi/2$ in the Shubnikov case.

\begin{figure}[t]
\centering
\includegraphics[width=0.5\textwidth]{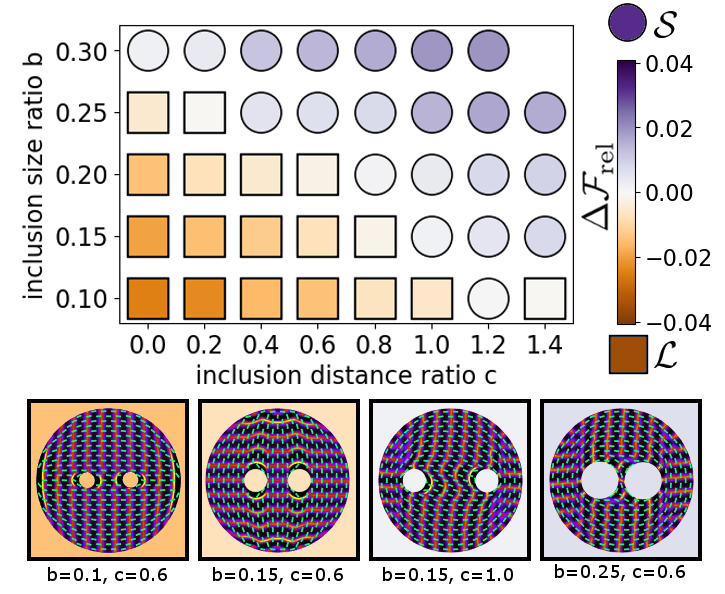}
\caption{
DFT state diagram in the two-holed-disk geometry indicating the stable laminar (square symbols) or Shubnikov (round symbols) states for different inclusion distance ratios $c$ and inclusion size ratios $b$.
The data for $c=0$, indicating the special case of annular confinement, are taken from Ref.~\cite{annulus}. The color denotes the relative free energy difference $\Delta \mathcal{F}_\text{rel}:=(\mathcal{F}_\text{0}^{(\mathcal{L})}-\mathcal{F}_\text{0}^{(\mathcal{S})})/\mathcal{F}_0$ between the optimal laminar $(\mathcal{L})$ and Shubnikov $(\mathcal{S})$ states, where $\mathcal{F}_0:=\min\{\mathcal{F}_\text{0}^{(\mathcal{L})},\mathcal{F}_\text{0}^{(\mathcal{S})}\}$ is the free energy of the overall optimal state. For $b=0.3$ and $c=1.4$ the inclusions are in contact with the outer wall, such that no Shubnikov state can exist.
The snapshots in the bottom panel depict four representative examples of optimal structures: laminar structures with tilt angles $\alpha=0$ or $\alpha=\pi/2$ (intermediate values of $\alpha$ are examined below in Fig.~\ref{fig_ANGLESDFTbuttonsF}) and Shubnikov structures without and with an inclusion tunnel.
}
\label{fig_PDDFTbuttons}
\end{figure}

Our results are compiled in Fig.~\ref{fig_PDDFTbuttons}.
We find that the laminar state is destabilized in favor of the Shubnikov state upon increasing the inclusion size ratio $b$, as in the special case $c=0$ of an annulus \cite{annulus}.
The laminar state is also destabilized upon increasing the spacing $c$ between the inclusions for a fixed value of $b$.
This behavior matches expectations, since the size of the effective inclusion increases when the two inclusions have a larger distance, such that Shubnikov structures,
characterized by layers spanning from the inclusions to the outer wall, become generally more favorable.
Since only the structure outside the convex hull of the inclusions is relevant for this first part of our discussion, it is not important whether or not the two inclusions are connected
(topological details arising from disconnected inclusions at $c>2b$ are discussed in Sec.~\ref{sec_BU_FS}).
However, we stress that, as soon as $c\geq 2-2b$, the inclusions overlap with the outer confining wall, such that it is no longer possible to fulfill the criterion to identify a Shubnikov state (cf.~the missing top-right state point in Fig.~\ref{fig_PDDFTbuttons}).
In these extreme cases, the confining geometry is, once again, simply connected and only deformed variants of a bridge state (laminar state without negatively charged defects)~\cite{annulus} exist, which are not our main interest here.
In the trivial bounding cases $c>2+2b$ or $b=0$ (not shown), the confinement simply reduces to a disk.

\begin{figure}[t]
\centering
\includegraphics[width=0.48\textwidth]{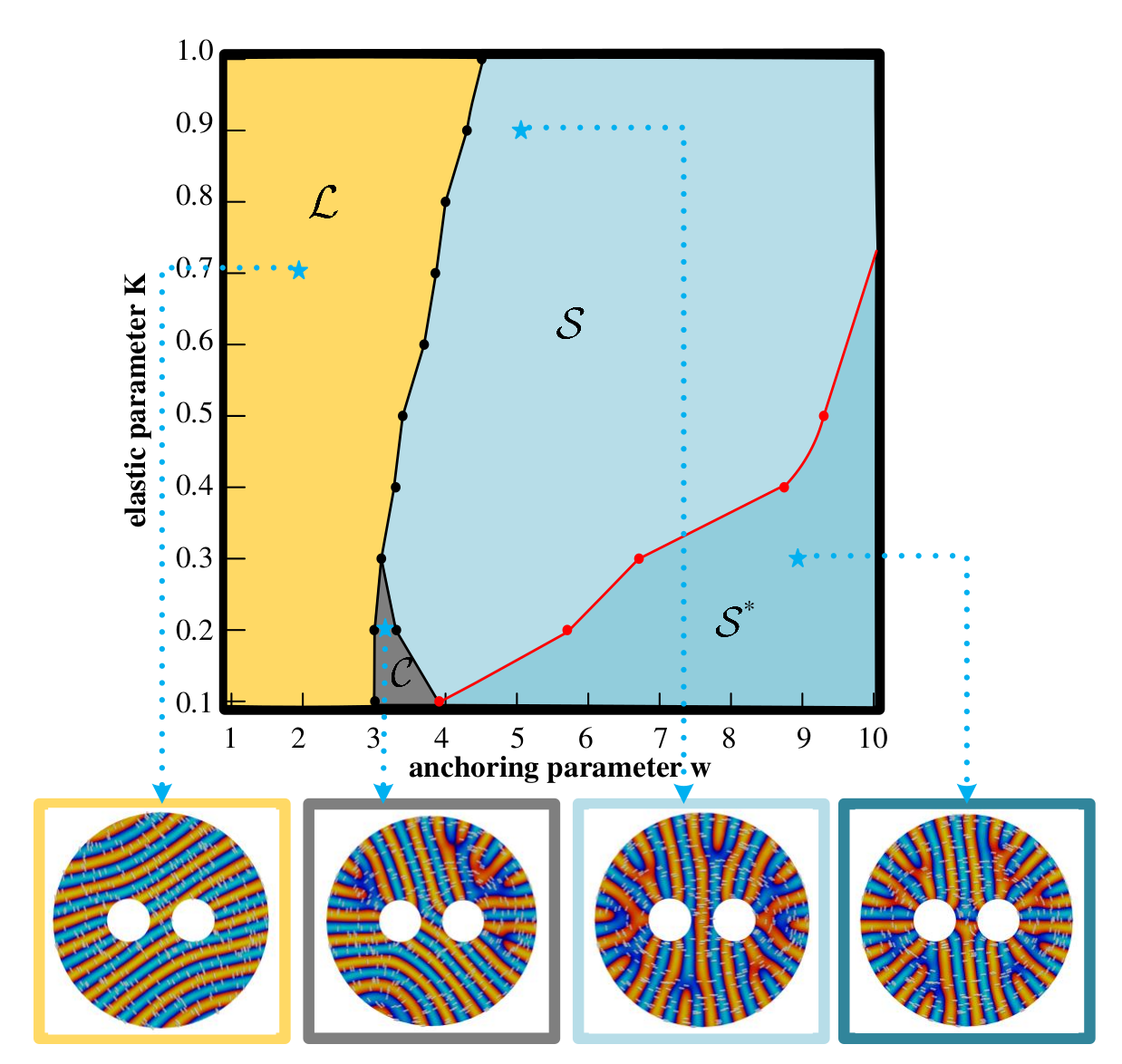}
\caption{State diagram from $\Qvec$-tensor theory in the two-holed-disk geometry for different values of the elastic parameter $K$ and the wall anchoring parameter $w$.
The inclusion size ratio $b=0.2$ and the inclusion distance ratio $c=0.6$ are kept fixed.
We distinguish between four different structures: laminar ($\mathcal{L}$), composite ($\mathcal{C}$) and Shubnikov without ($\mathcal{S}$) and with an inclusion tunnel ($\mathcal{S}^*$).
The bottom panel depicts one representative snapshot for each case (parameters according to the connected stars).
}
\label{fig_PDDFTbuttonsLdG}
\end{figure}

In general, the laminar-Shubnikov transition is driven by the tendency of the system to achieve an optimal balance between satisfying the external constraints of the confining geometry (since the rods preferably align parallel to the wall)
and maintaining the intrinsic smectic structure.
This results in a trade-off between deformations, as dominant in the Shubnikov structures, and topological defects, governing the laminar structures.

To better characterize this competition,
we employ the smectic $\Qvec$-tensor theory to examine how the structural transitions can be induced by tuning the
elastic behavior and the strength of the tangential wall alignment,
determined by the parameters $K$ and $w$, respectively.
To this end, we fix the inclusion size ratio $b=0.2$ and the inclusion distance ratio $c=0.6$, so as to take values close to the laminar-Shubnikov transition predicted by DFT in Fig.\ \ref{fig_PDDFTbuttons}. The state diagram from $\Qvec$-tensor theory in Fig.\ \ref{fig_PDDFTbuttonsLdG} confirms the expectation that Shubnikov states are stabilized upon imposing stronger anchoring conditions (i.e., larger $w$) to minimize the number of defects.
In particular, the observed laminar states are typically characterized by a single domain with the defects appearing through a misalignment at the walls (as also frequently observed in DFT).
Moreover, we see that the laminar state is generally stabilized upon increasing $K$ and thus the bending rigidity of the layers.
For smaller values of $K<0.3$, composite structures are also found and the laminar state becomes compatible with strongly deformed layers.
Such highly elastic behavior is, however, rather atypical in the context of hard rods.
These results demonstrate a reassuring consistency between the DFT and smectic $\Qvec$-tensor results, even without extensive tuning of the other parameters of the $\Qvec$-tensor model.

Returning to microscopic DFT structures, we can make more precise statements regarding the stability of the generalized laminar and Shubnikov states, by comparing the examples shown at the bottom of Fig.~\ref{fig_PDDFTbuttons}.
First of all, both the number of layers in the central domain and their orientation in the optimal laminar state (under the symmetry constraints imposed so far) strongly depend on the geometrical parameters. This suggests that by allowing for different values of the tilt angle $\alpha$ we should find states with an even smaller free energy, which will be studied in Sec.~\ref{sec_BU_OR}. Second, and most importantly, we notice for $\alpha=\pi/2$ that the structural differences between states classified as laminar or Shubnikov become less pronounced upon increasing the inclusion distance $c$, due to the larger size of the central domain.
This intuitively explains the destabilization of the laminar state for increasing $c$:
the number of laminar layers between the inclusions and outer wall decreases, which brings the two defects closer to annihilation, while the extreme case of zero laminar layers eventually corresponds to a Shubnikov structure.
Third, for $b=0.1$, we even observe in Fig.~\ref{fig_PDDFTbuttons} a particular example of a re-entrant stable laminar state at $c=1.4$, in which the spacing between the inclusions and outer wall allows for all layers being parallel.
Finally, we expect that, within a small range of parameters, there exists a stable intermediate composite state \cite{annulus}, as in the illustrations in the middle panel of Fig.~\ref{fig_structures}.

\begin{figure}[t]
\centering
\includegraphics[width=0.5\textwidth]{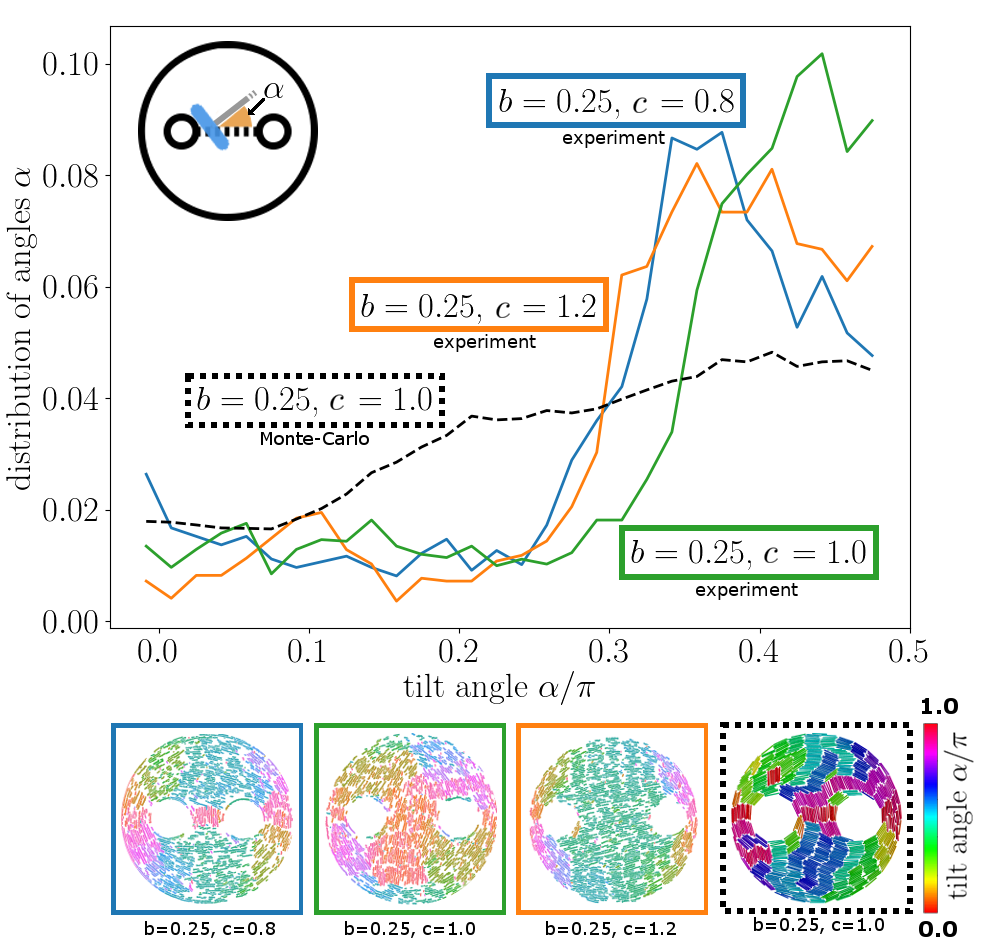}
\caption{Global orientational distribution in the two-holed-disk geometry for the inclusion size ratio $b=0.25$. As illustrated at the top right, the relative frequency of individual rod orientations (blue) reflects the preferred tilt angle $\alpha$ of the layers (gray line) relative to the axis that connects the two inclusions (dotted line), see also Fig.~\ref{fig_structures}.
We compare experimental data for different inclusion distance ratios $c\approx0.8$, $c\approx1.0$ and $c\approx1.2$, averaged over 21, 24 and 15 available structures, respectively, and Monte-Carlo data for $c=1.0$, sampled from $1800$ independent simulation runs.
For symmetry reasons we map orientations with $\alpha > \pi/2$ onto $\pi-\alpha$ and only consider $0\leq\alpha\leq\pi/2$.
The bottom panel depicts one particular snapshot for each case, where the rods are colored according to their orientation.
}
\label{fig_ANGLESDFTbuttons}
\end{figure}

\begin{figure}[t]
\centering
\includegraphics[width=0.4975\textwidth]{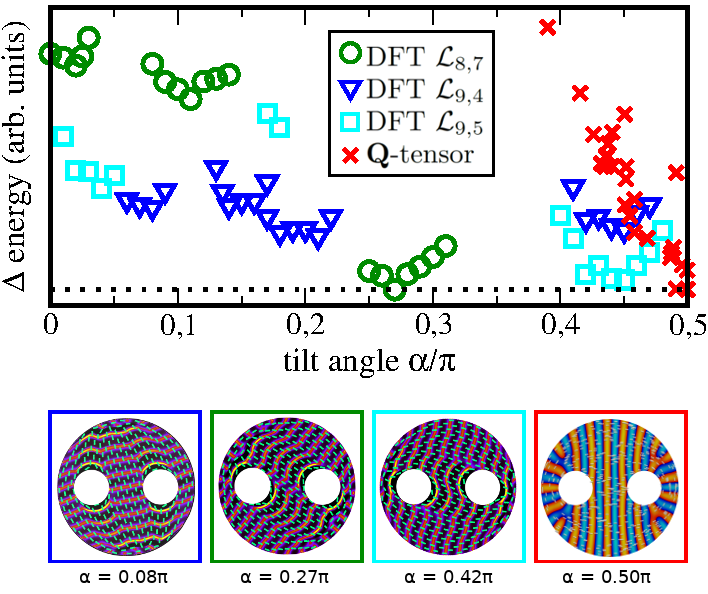}
\caption{Energy landscape for different structures depending on the tilt angle $\alpha$ of the central domain in the two-holed-disk geometry for the inclusion size ratio $b=0.25$ and the inclusion distance ratio $c=1.0$.
According to the legend, we compare different laminar DFT structures $\mathcal{L}_{ab}$, where the indices $a$ and $b$ denote the number of layers in the central domain and perpendicular to it, respectively, and several minimizers of the energy functional from $\Qvec$-tensor theory with $K=1$ and $w=5$.
Exemplary snapshots are shown in the bottom panel.
As only the energy difference is relevant for the stability, the vertical axis depicts the rescaled difference to the global minimum (indicated by the dotted line), calculated separately for DFT and $\Qvec$-tensor results in arbitrary units.
Since the minima for large angles $\alpha$ are generally deeper, it is more likely to find such structures, consistent with the observation in Fig.~\ref{fig_ANGLESDFTbuttons}.
}
\label{fig_ANGLESDFTbuttonsF}
\end{figure}

\subsubsection{General particle-based observations \label{sec_BU_OBS}}

In our colloidal experiments, we focus on a few selected sets of geometrical parameters.
Qualitatively inspecting our snapshots for $b\approx0.25$ and $0.6<c<1.2$, we arrive at the following general picture.
We predominantly observe the Shubnikov state, in agreement with the DFT prediction.
All of these Shubnikov structures possess large tilt angles $\alpha>\pi/4$ of the central domain.
Recalling the discussion in Sec.~\ref{sec_classification}, such an alignment allows for a larger number of straight layers in the central domain between the inclusions. Quite remarkably, however, only one of our $104$ inspected structures depicts a nearly laminar state (see the second snapshot in Fig.~\ref{fig_ANGLESDFTbuttons}), while only three of them can be clearly identified as composite states.
In all these cases, the laminar parts of the structure possess a small tilt angle $\alpha<\pi/4$.
 To quantify the tilt-angle statistics, we measure in Fig.~\ref{fig_ANGLESDFTbuttons} the global orientational distribution of all rods, averaged over all cavities with comparable geometry.
In accordance with the typical orientation $\alpha$ of the central domain, we find that the most frequent angles are close to $\pi/2$, where the exact location of this peak appears to  depend on the inclusion distance.

Overall, the suppression of the stability of laminar states, upon increasing the distance between the inclusions, appears to be even more pronounced in the experiments than predicted theoretically in Fig.~\ref{fig_PDDFTbuttons}.
This observation can be explained by the typically lower number of parallel layers in the experiment compared to the most stable DFT solution \cite{annulus} in combination with the preference of the rods to align in a central domain at large tilt angles.
To understand this, consider, for instance, the experimental laminar structure depicted in Fig.~\ref{fig_ANGLESDFTbuttons} with $b=0.25$ and $c=1.0$.
Now imagine, instead, the inclusions placed over the top and bottom grain boundary. This would both reduce the defect region and classify the structure as a Shubnikov state with a significantly increased tilt angle $\alpha$,
intuitively explaining our predominant observations of large tilt angles and Shubnikov structures.

Our Monte-Carlo simulations of hard rods carried out for $b=0.25$ and $c=1.0$ confirm the basic experimental observations that nearly all identified structures reflect stretched Shubnikov states and that large tilt angles are favored. We depict the global orientational distribution and the typical snapshot at the bottom-right of Fig.~\ref{fig_ANGLESDFTbuttons}.
Moreover, our particle-resolved simulations allow us to further explore smaller inclusion sizes than those realized experimentally (not shown).
As expected from the discussion above, there is still a high probability to observe Shubnikov structures for $b=0.1$, while the laminar state becomes dominant for $b=0.05$.

\subsubsection{Orientation of the central domain \label{sec_BU_OR}}

Our experimental and Monte-Carlo results suggest that the assumption, made in Sec.~\ref{sec_BU_LS} for DFT, of imposing smectic structures with the same symmetry as the confining geometry is not justified in general.
 While the orientational distribution in Fig.~\ref{fig_ANGLESDFTbuttons} generally suggests that large tilt angles $\alpha>\pi/4$ are most likely, we also notice that the maximum is not always located at the extreme value $\alpha=\pi/2$. Even more so, we expect that the geometrical constraints on laminar structures, arising from the competition of the preferred layer spacing with both the distance between the two inclusions and the distance from each inclusion to the outer wall, can be efficiently relaxed by aligning the central domain along characteristic tilt angles. A first evidence for this prediction stems from the DFT results in Fig.~\ref{fig_PDDFTbuttons}, where the tilt angle of the optimal laminar structure (given the constraint to either $\alpha=0$ or $\alpha=\pi/2$) strongly depends on the particular geometry (contrast the two depicted laminar structures).

To learn more about the preferred tilt angle, we compare in Fig.~\ref{fig_ANGLESDFTbuttonsF} the energy of different states as a function of $\alpha$ for a fixed geometry with $b=0.25$ and $c=1.0$.
In the smectic $\Qvec$-tensor theory, we only find solutions with large tilt angles
$\alpha>0.4\pi$ for the intrinsic parameters $K=1$ and $w=5$, which demonstrates the instability of structures with smaller $\alpha$ under these conditions.
The corresponding free energy decreases with increasing tilt angle, such that the global minimum is found for $\alpha\simeq0.5\pi$, which is in principal agreement with the statistics from experiment and Monte-Carlo simulation.

To systematically study the tilt-angle dependence in DFT, we restrict ourselves to laminar states. We choose three representative template structures with a well-defined numbers of layers both in the central domain and perpendicular to it (cf.~the example structures shown at the bottom of Fig.~\ref{fig_ANGLESDFTbuttonsF}).
 By doing so, all structures generated by imposing different tilt angles remain comparable among each other.
For the parameters $b=0.25$ and $c=1.0$, we find that structures with two layers interrupted by each inclusion are generally favorable.
We further focus in each case on three typical ranges of the tilt angle, such that there are (with increasing $\alpha$) three, two or one
laminar layers between one inclusion and the outer wall, respectively.
These values of $\alpha$ depend on whether the central point of the geometry is occupied by a layer (central domain with nine layers in total) or by the void space in between two layers (central domain with eight layers in total). The corresponding free energy landscapes shown in Fig.~\ref{fig_ANGLESDFTbuttonsF} reveal that the most stable structures correspond to the minima in the range of tilt angles with the largest values.
This reflects the intuition that the two inclusions are preferably located close to (or even on) the edge of the central domain and not in its center, such that the extent of deformations of the central smectic layers is reduced.
The existence of distinct local free energy minima in Fig.~\ref{fig_ANGLESDFTbuttonsF} explains the nonmonotonic and geometry-dependent experimental distributions in Fig.~\ref{fig_ANGLESDFTbuttons}.

\begin{figure}[t]
\centering
\includegraphics[width=1.0\linewidth]{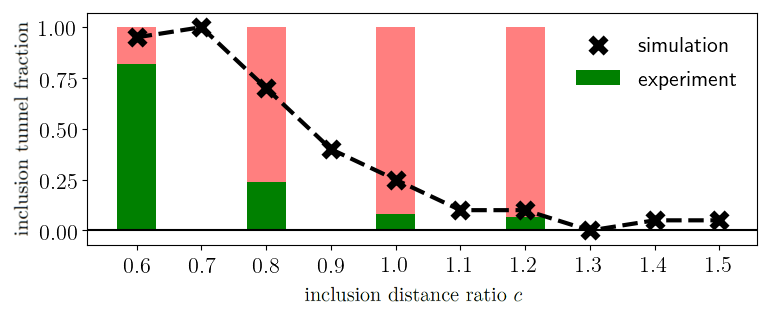}
\caption{
Relative frequencies of structures with an inclusion tunnel (cf.\ the second row in Fig.~\ref{fig_structures}) in the two-holed-disk geometry for the inclusion size ratio $b=0.25$.
We compare experimental data (green bars) for different inclusion distance ratios $c\approx0.6$, $c\approx0.8$, $c\approx1.0$ and $c\approx1.2$, averaged over 44, 21, 24 and 15 available structures, respectively, and Monte-Carlo data (black crosses) averaged over 20 simulations for each selected $c$.
The dotted line serves as a guide to the eye, illustrating how the fraction of inclusion tunnels decreases with increasing inclusion distance.
Regarding the occurrence of structures with inclusion tunnel in DFT and $\Qvec$-tensor theory, please refer to the bottom-right snapshot in Fig.~\ref{fig_PDDFTbuttons} and the state diagram in Fig.~\ref{fig_PDDFTbuttonsLdG}, respectively.}
\label{fig_P_inclbridge}
\end{figure}

\begin{figure*}[t]
\centering
\includegraphics[width=1.0\linewidth]{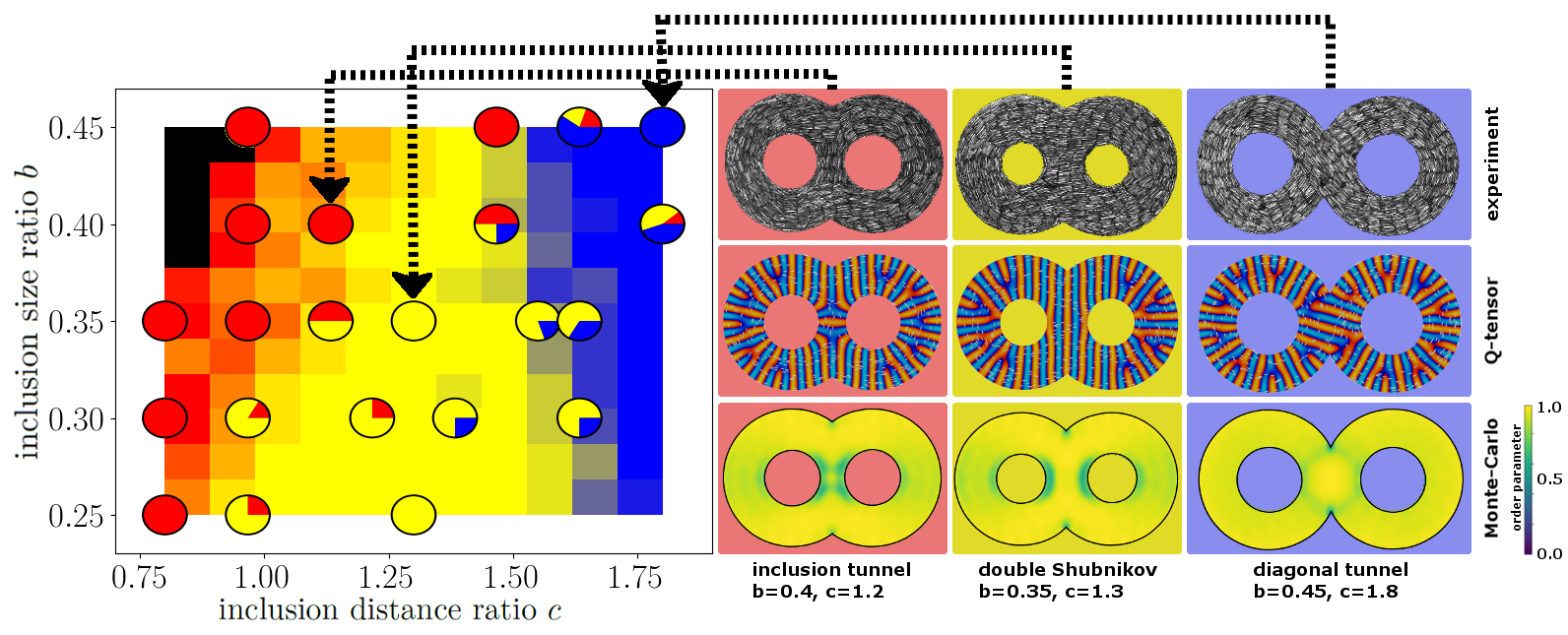}
\caption{
Structures in the double-annulus geometry for different inclusion size ratios $b$ and inclusion distance ratios $c$.
While we generally observe layering of the Shubnikov type in the annular arcs for the parameters considered, we distinguish three states by the layer arrangement in the intersection region:
 inclusion tunnel (red), double-Shubnikov state (yellow) and diagonal tunnel (blue).
Left panel: state diagram indicating the relative frequencies of the three structures between the two inclusions.
The experimental and Monte-Carlo results are represented by pie charts and background pixels with proportional color mixing, respectively.
For $c\lesssim2b$ no rods fit in between the inclusions and there is no distinction (black pixels).
Right panel: observed structures for three pairs of $b$ and $c$ corresponding to the state points indicated by the arrows.
We depict experimental snapshots (top), solution profiles of the $\Qvec$-tensor model with
for $K=1.0$ and $w=10$ (middle) and the orientational order parameter $\Sop\ofr$ averaged over $10^3$ independent Monte-Carlo simulation runs per parameter pair, revealing the typical location of the topological defects through the darker shades (bottom).
}
\label{fig_FAinclusions}
\end{figure*}

\subsubsection{Fine structure between the inclusions \label{sec_BU_FS}}

Having understood the large-scale layering behavior of the central domain, we now investigate the structure inside the convex hull of the inclusions in more detail.
For $c>2b$, the two inclusions are disconnected and we anticipate two additional (compared to a single or two overlapping inclusions) topological defects with a negative charge.
As generally described in Sec.~\ref{sec_classification}, there are two possible scenarios,
related to how the smectic layers between the inclusions align.
The first possibility, which has been silently implied so far when discussing our large-scale results in the previous sections, is that the central domain and the layers between the inclusions align with each other, compare the first three rows of Fig.~\ref{fig_structures}.
To be more specific,
we can conclude that the layers need to fill the space between the inclusions in an entropically convenient way is probably one of the main driving forces that determines the geometry-dependent tilt angle of the central domain as a whole.

The second possibility of alignment between the inclusions is an inclusion tunnel, compare the last row of Fig.~\ref{fig_structures}.
This structure is defined by one or more smectic layers spanning between the two inclusions,  irrespective of the orientation of the central domain. The driving force behind the formation of an inclusion tunnel is the adherence to the preferred wall alignment which comes at the cost of a larger grain boundary within the system.
This is nicely reflected by additionally differentiating in the state diagram from $\Qvec$-tensor theory, as in Fig.~\ref{fig_PDDFTbuttonsLdG}, between Shubnikov states with and without an inclusion tunnel.
It is apparent from the state diagram that structures with an inclusion tunnel stabilize upon increasing the anchoring parameter $w$ and decreasing the elastic parameter $K$.

In our DFT study, we find that structures with the central domain interrupted by an inclusion tunnel are almost always less stable than comparable ones with a continuous central domain, for both laminar and Shubnikov structures alike.
The fact that the central domain tends to tilt, renders such an inclusion tunnel even less favorable due to the general preference of hard rods to meet at a grain boundary with nearly perpendicular orientations, instead of an oblique alignment. An inclusion tunnel only becomes energetically favorable for extremely small distances between the surfaces of the two inclusions, of about one rod length or less, as e.g.~for $b=0.25$ and $c=0.6$, compare the fourth structure shown at the bottom of Fig.~\ref{fig_PDDFTbuttons}.

In practice, however, it is much more likely to observe these inclusion tunnels as a result of the equilibration protocol.
More specifically, in our experiments and Monte-Carlo simulations, the growth of an inclusion tunnel can be triggered by small domains aligning with the inclusion at an early stage.
Hence, such structures are observed with a noticeable probability, even for relatively large $c$, as verified in Fig.~\ref{fig_P_inclbridge}.

\subsection{Double annulus}
\noindent
We have seen in Sec.~\ref{sec_BU} that
the inclusion distance ratio $c$ and, therefore, the minimal distance from the inclusions to the outer wall is an important criterion which determines the globally observed state in the two-holed-disk geometry.
The smectic structure between the inclusions then largely follows the alignment of the central domain, while inclusion tunnels are only rarely observed.

Now we focus on the double-annulus geometry, illustrated at the bottom of Fig.~\ref{fig_concept}, for which a larger range $b<c<2$ of inclusion distance ratios $c$ can be examined without changing the Euler characteristic $\chi=-1$.
Since the shortest distance from any point on the outer wall to one of the inclusions remains the same for all $c$, the smectic structure in the two annular arcs is largely determined by the inclusion size ratio $b$ alone and can thus be well understood by taking cues from the state diagram in annular confinement \cite{annulus}.
This gives us a better control of how the central smectic layers in the intersection region of the two annular halves respond to changes of the inclusion distance compared to the single circular outer wall of the two-holed-disk geometry.
We are thus primarily interested in the question of how the structure between the two inclusions of the double annulus is determined by the geometrical parameters $b$ and $c$, as we focus on inclusion size ratios $b\geq 0.25$ which predominantly give rise to generalized Shubnikov structures in the annular arcs.

Our state diagram, compiled from experiments and particle-resolved Monte-Carlo simulations, is shown in the left panel of Fig.~\ref{fig_FAinclusions}.
Both methods consistently predict three different types of structures, shown in the right panel.
First, for relatively large and nearby inclusions, we typically observe an inclusion tunnel, similar to the two-holed-disk geometry (cf.\ Sec.~\ref{sec_BU_FS}).
Second, for relatively small and distant inclusions, we typically observe a structure with a large central domain of vertical layers, which is similar to the $\alpha=\pi/2$ alignment in the two-holed-disk geometry (cf.\ Sec.~\ref{sec_BU_OR}).
As mentioned in the previous paragraph this extreme tilt angle is favored here due to the broken rotational symmetry and the non-convex shape of the outer wall.
We refer to such a structure as the double Shubnikov state, as there are no grain boundaries (the two $Q=-1/2$ defects are mostly due to misalignment at the inclusions).
Third, for relatively large and distant inclusions, we typically observe a structure which is characterized by both a large tilted central domain and grain boundaries.
The tilt angle is again roughly set by the geometry, such that the orientation of the rods follows an infinity symbol.
This diagonal-tunnel state possesses no analog in the two-holed-disk geometry.
To corroborate these observations, we also evaluated our $\Qvec$-tensor theory for representative pairs of parameters and find consistent minimizers, shown in the right panel of  Fig.~\ref{fig_FAinclusions}. Moreover, the exemplary double-Shubnikov structures shown in Fig.~\ref{fig_structures} using all four methods are in close agreement.

To further highlight the topological distinction between the three different structures observed in the double-annulus geometry, we additionally show in the right panel of Fig.~\ref{fig_FAinclusions} Monte-Carlo results for the local order parameter field $\Sop\ofr$, sampled as an average from $10^3$ independent simulation runs.
Due to the averaging, we obtain in each case a characteristic pattern, which possesses the same symmetry as the confinement.
The inclusion tunnel is characterized by its orthogonal alignment relative to the nearby layers and therefore a large degree of orientational frustration between the inclusions. In the double Shubnikov state, the region between the inclusion largely aligns with the central domain and the orientational frustration is manifest only close to the inclusions (usually due to small domains of a few rods).
Finally, for the diagonal tunnel, it is clearly visible that the grain boundaries are located at the edges of the central crossing of the annular arcs.

 \section{Summary and conclusions \label{sec_conclusions}}

In this work, we investigate smectic states, confined to complex geometries, illustrated in Fig.~\ref{fig_concept}, with two circular inclusions (interior boundaries) by means of colloidal experiments, Monte-Carlo simulations, density functional theory (DFT) and smectic $\Qvec$-tensor theory.
Our four approaches consistently predict the main structural features, as exemplified in Fig.~\ref{fig_overview}.
All observed and expected structures are compiled in Fig.~\ref{fig_structures}.

For large inclusions (or strong wall anchoring), the layers arrange into a generalized Shubnikov state, characterized by an overall perpendicular alignment of layers (or parallel alignment of rod-like particles) at the outer wall, which minimizes the number of topological defects.
This is observed in both the two-holed-disk geometry (see the circular data points in Fig.~\ref{fig_PDDFTbuttons} and the bottom-right and central regions (both shades of blue) in Fig.~\ref{fig_PDDFTbuttonsLdG}), where a stretched Shubnikov state also stabilizes for increasing inclusion distance, and the double-annulus geometry (see all data in Fig.~\ref{fig_FAinclusions}).
On the contrary, for small inclusions (or weak wall anchoring), the layers arrange into a generalized laminar state characterized by two $Q=-1/2$ defects at either of the two inclusions and two $Q=+1/2$ defects close to the outer wall.
This is explicitly observed in the two-holed-disk geometry (see the quadratic data points in Fig.~\ref{fig_PDDFTbuttons} and the leftmost region (yellow) in Fig.~\ref{fig_PDDFTbuttonsLdG}) but we expect the same upon further decreasing the inclusion size the double-annulus geometry.

If the two inclusions are sufficiently close to each other, we observe an inclusion tunnel in both the two-holed-disk geometry (see the bottom-right structure in Fig.~\ref{fig_PDDFTbuttons}, dark blue color in Fig.~\ref{fig_PDDFTbuttonsLdG} and the statistics in Fig.~\ref{fig_P_inclbridge}) and the double-annulus geometry (see the data with red color in Fig.~\ref{fig_FAinclusions}).
This structure forms an isolated domain between the two inclusions and two grain boundaries, irrespective of the global state.
More distant inclusions allow for the layers to align in a larger central domain at the cost of misalignment at the inclusions.
In fact, in the two-holed-disk geometry, this relative alignment of the central layers to the axis connecting the two inclusions is characterized by large tilt angles $\alpha\simeq\pi/2$ (see Figs.~\ref{fig_ANGLESDFTbuttons} and~\ref{fig_ANGLESDFTbuttonsF}).
In the double-annulus geometry, we further distinguish between two cases (identified here for generalized Shubnikov states).
The double Shubnikov structure possesses a large central domain which extends over all four ends of the geometry's central junction at a fixed tilt angle $\alpha\approx\pi/2$ (see the data with yellow color in Fig.~\ref{fig_FAinclusions}), while for even larger inclusion distances, we observe a diagonal tunnel, characterized by two grain boundaries at two opposing ends of the central junction and a tilt angle $0<\alpha<\pi/2$ dictated by the geometry (see the data with yellow color in Fig.~\ref{fig_FAinclusions}).

Our study represents a first step towards the study of liquid crystals confined to topologically highly complex environments such as random porous media \cite{guegan2006evidence,scholz2012permeability,coll_transp,avendano2016assembly} or  arrays of obstacles \cite{coll_transp}.
Our complementary approaches can, in principle, be applied to any kind of confinement \cite{monderkamp2021topology,slitpores,real_defch,garlea2019colloidal}.
This applies in particular also to systems in three dimensions to which our experimental, computational and theoretical methods, as well as, our topological concepts can be generalized \cite{monderkamp2022topological,wittmann2016,xia-2021-article}.
Another generalization is to proceed towards more complex particle shapes and interactions such as hard polygons \cite{gantapara2015novel,avendano2016assembly}, non-convex  \cite{hernandez2007colloidal,niori1996distinct,heppke2000novel,dingemans2000non,ros2005banana}, or chiral particles \cite{monderkamp2023network,kamien2001order,harris1997microscopic,harris1999molecular,pollard2019point,meyer1977ferroelectric,dierking2014chiral,hoell2016colloidal}.
Finally it bears mentioning that many bacteria have rod-like shapes \cite{wensink2012meso,van2018recent,allen2018bacterialreview,wittmann2022mechano}
and are living on two-dimensional substrates, where they can be easily be put in confinement \cite{volfson2008biomechanical,wioland2013confinement}.
Bacterial colonies can approach high densities, where smectic layering is expected \cite{boyer2011buckling,you2018geometry,langeslay2022}, such that our work may have important consequences for the structure in dense biofilms.

 One compelling open question concerns the existence of similar structures and the applicability of our topological methods for smectic phases of molecular liquid crystals, a central aspect of experimental liquid crystal research \cite{10.1021/acs.nanolett.9b04347,10.1073/pnas.2000849117,10.1021/acs.langmuir.5b02508,10.1002/adom.201500153,originalPointsToLines,10.1039/d0sm01112f,10.1021/acs.langmuir.7b03351,10.1007/s00396-010-2367-7,ma13173761,Gim2017}.
While our hard-rod model is specifically designed to mimic our colloidal experiments, the analogous observations by means of $\Qvec$-tensor theory leave us optimistic that this gap can be bridged in future work on molecular systems.
Regarding the topological analysis, it might prove fruitful to focus on the smectic layers \cite{aharoni2017composite,machon2019} instead of the orientational director when studying molecular liquid crystals, for which it is no longer possible to achieve a particle resolution.

 \acknowledgments

RW and HL acknowledge support by the Deutsche Forschungsgemein\-schaft (DFG) through the SPP 2265, under grant numbers WI 5527/1-1 (RW) and LO 418/25-1 (HL).
The work of JX is supported by the National Natural Science Foundation of China (No.~12201636) and the Research Fund of National University of Defense Technology [grant number ZK22-37].
The work of PEF is supported by the Engineering and Physical Sciences Research Council [grant numbers EP/R029423/1 and EP/W026163/1].
PEF and RW acknowledge the support of the Banff International Research Station (workshop 22w5159).

 \appendix

\section{Details on the methods \label{app_methods}}

\subsection{Experiment \label{app_methods_exp}}
The experimental methods follow from our previous work \cite{CollLQinSqConf,annulus}.
In short, home-synthesized silica rods \cite{kuijk2011synthesis} suspended in a 1mM NaCl water solution form sedimentation-diffusion equilibrium into a cylinder-shaped reservoir glued to a glass coverslip (see Fig.~1a in Ref.~\cite{annulus}).
Confinement cavities, as shown in the first row of Fig.~\ref{fig_overview}, are printed on the coverslip using polydimethylsiloxane (PDMS) mold and Norland Optical Adhesive glue. PDMS molds are made using standard soft lithography technique.

Particles have average effective length of $5.3\mu$m, aspect ratio of $10.6$ and gravitational length of $0.8\mu$m.
After insertion they start forming a concentration gradient along the direction of gravity.
At the bottom, inside the cavities, we successively observe the formation of isotropic, nematic and finally smectic phases.
The total amount of particles is chosen such that there is no crystalline state. The smectic structures are left to equilibrate for at least 12 hours.

Experimental snapshots capture the rods in direct vicinity of the bottom wall of the cavity that, paired with gravity, imposes a quasi-two dimensional confinement. We record images by mean of confocal microscopy with a Zeiss LSM Exciter 5 microscope and a 63x Zeiss Plan Apo Chromat objective. We collect  scattered light to form images as this batch of rods is not fluorescent.

A custom python script is used to segment single rods and detect position and orientation (a Wolfram Mathematica script is already available \cite{annulus}). The specific python script used to process the snapshots of Fig.~\ref{fig_ANGLESDFTbuttons} is provided along with an experimental snapshot as supplementary material.

\subsection{Monte-Carlo simulations \label{app_methods_sim}}

With the help of canonical Monte-Carlo simulations we generate equilibrium states for liquid crystals composed of hard rods at bulk smectic area fraction $\eta_2 = 0.725$. The rods are modeled as discorectangles with aspect ratio $p=L/D = 16.5$, where $L$ denotes the length and $D$ the width of the particles. The $k$th rod is parametrized by a line segment $\mathbf{a}_k = \mathbf{r}_k + \alpha_k \mathbf{\hat{u}}_k$, with position $\mathbf{r}_k$, normalized orientation $\mathbf{\hat{u}}_k$ and $\left | \alpha_k \right | < L/2$. All points within the area of the rod are characterized by $\{ \mathbf{x} \in \mathbb{R}^2| \left\|  \mathbf{x} - \mathbf{a}_k \right\| \leq D/2 \}$ such that the standard hard-core repulsion between a pair of rods $i$, $j$ can be defined by
\begin{align}
U(\mathbf{r}_i,\mathbf{r}_j,\mathbf{\hat{u}}_i,\mathbf{\hat{u}}_j) =
\begin{cases}
\infty & \text{ for } d_{i,j} \leq D\,, \\
0 & \text{ for } d_{i,j} > D\,,
\end{cases}
\end{align}
where
\begin{equation}
d_{i,j} = \min_{\left | \alpha,\beta \right | < \frac{L}{2}} \left \| \mathbf{r}_i
+ \alpha \mathbf{\hat{u}}_i - ( \mathbf{r}_j + \beta \mathbf{\hat{u}}_j ) \right \|
\end{equation}
corresponds to the smallest distance between the opposing line segments \cite{overl}.

The interaction of the rods with the walls is modeled by considering the rods as three virtual point particles at $\mathbf{r}_k + \gamma  \mathbf{\hat{u}}_k$, $\gamma \in \{ -L/2,0,L/2 \}$.
The wall potential reads as
\begin{align}
\label{eq_wallpot}
V(x) =
\begin{cases}
\Phi(x_{0}) +\Phi'(x_{0})(x-x_0) & \text{ for } x \leq x_{0}\,, \\
\Phi(x) & \text{ for } x_{0} > x\,.
\end{cases}
\end{align}
Here, $\left | x \right |$ denotes the minimal perpendicular distance from either
of the two points to the wall and $x>0$ corresponds to the inside of the cavity.
The cut-off point, below which $V(x)$ is linear, is chosen as $x_0=0.5D$.
For $\Phi(x)$, we choose the standard 12-6-Weeks-Chandler-Andersen-potential \cite{WCA}
\begin{equation}
\Phi(x) =
\begin{cases}
4\epsilon \left [ \left ( \frac{D}{x} \right )^{12} - \left ( \frac{D}{x} \right )^{6} \right ] + \epsilon & \text{ for } x \leq 2^{\frac{1}{6}} D\,, \\
0 & \text{ for } x > 2^{\frac{1}{6}} D
\end{cases}
\end{equation}
with $\epsilon=10k_\text{B}T$, with the Boltzmann constant $k_\text{B}$ and temperature $T$.
The potential landscapes to model the two-holed-disk and double-annulus geometries can be expressed as combination of circular well and obstacles. The outer radius of the cavity is chosen as $\Rout = 6L$.

To obtain the equilibrated configurations, we initialize the system at a dilute area fraction $\eta_0 = 0.01$. We subsequently compress the system, by rescaling the cavity, at a compression rate of  $\Delta \eta_1= 3.50 \times 10^{-7}$ per Monte-Carlo cycle to an intermediate area fraction just below the bulk isotropic-nematic phase transition. In a second stage, we compress the system with $\Delta \eta_1= 7.33 \times 10^{-8}$ per Monte-Carlo cycle to the final area fraction $\eta_2 = 0.725$. The area fraction is given by the fraction of the sum of the individual volumes of the rods $\Vrod$ to the total volume of the cavity $\Vcav$.
Since the final area fraction and the final volume are fixed variables, by the geometric parameters $b$, $c$ (see Fig.~\ref{fig_concept}) and $\Rout$ in terms of the particle size, the particle number $N$ remains a free parameter that is determined at the start of the simulation via the relation
\begin{equation}
    \eta = \frac{N\Vrod}{\Vcav} = \frac{N}{\Vcav} \left( \frac{\pi D^2}{4} + D L\right).
    \label{eq_eta}
\end{equation}
The typical values for $N$ we investigate are on the scale of several thousand.
Typical snapshots in the two geometries are shown in the second row of Fig.~\ref{fig_overview}.

\subsection{Density functional theory (DFT) \label{app_methods_DFT}}

Classical density functional theory (DFT) \cite{Evans1979} allows us to predict the structure of anisotropic fluids in an external potential $V_\text{ext}(\bvec{r},\phi)$ by calculating the equilibrium density profile $\rho(\bvec{r},\phi)$ from a variational principle, where $\bvec{r}$ denotes the center-of-mass position and $\phi$ the particle orientation.
This is achieved by minimizing the grand potential functional
\begin{align}
 \Omega[\rho]=\mathcal{F}[\rho]+\!\int\!\mathrm{d}\bvec{r}\!\int_0^{2\pi}\!\frac{\mathrm{d}\phi}{2\pi}\,\rho(\bvec{r},\phi)(V_\text{ext}(\bvec{r},\phi)-\mu)\,,
\end{align}
at given chemical potential $\mu$ by iterating the Euler-Lagrange equation $\delta\Omega[\rho]/\delta\rho(\bvec{r},\phi)=0$, where $\mathcal{F}[\rho]$ is the intrinsic Helmholtz free energy functional.
The solution density profile $\rho(\bvec{r},\phi)$ for a given initial guess is given by a local minimum of the grand potential $\Omega$.
Here, we minimize under the constraint of a fixed total particle number $\int\!\mathrm{d}\bvec{r}\!\int_0^{2\pi}\!\frac{\mathrm{d}\phi}{2\pi}\,\rho(\bvec{r},\phi)$, to obtain local minima of the Helmholtz free energy $\mathcal{F}$.

For an explicit calculation, we need to specify the Helmholtz free energy functional $\mathcal{F}[\rho]=\mathcal{F}_\text{id}[\rho]+\mathcal{F}_\text{ex}[\rho]$,
which is conveniently split into an exactly known ideal part
\begin{align}\label{eq_freeenergy}
\beta\mathcal{F}_\text{id}[\rho]=\!\int\!\mathrm{d}\bvec{r}\!\int_0^{2\pi}\!\frac{\mathrm{d}\phi}{2\pi}\,\rho(\bvec{r},\phi) \left(\ln (\rho(\bvec{r},\phi)\Lambda^2)-1\right)
\end{align}
and an excess part $\mathcal{F}_\text{ex}[\rho]$.
The irrelevant thermal wave length $\Lambda$ is set to unity the inverse temperature $\beta:=(k_\text{B}T)^{-1}$ is just a scaling factor.
The excess free energy is based on fundamental measure theory \cite{Rosenfeld1989,roth2010fundamental,roth2012} for anisotropic hard particles in two dimensions \cite{PD_Rene,annulus},
expressing the functional $\mathcal{F}_\text{ex}[\rho]$ as a function of weighted densities
\begin{equation}
 \label{eq_weighdenorient}
  n_{\nu}(\bvec{r}) = \!\int\!\mathrm{d}\bvec{r}_1\!\int_0^{2\pi}\!\frac{\mathrm{d}\phi}{2\pi}\,\rho(\bvec{r}_1,\phi)\, \omega^{(\nu)}(\bvec{r}-\bvec{r}_1,\phi) \,.
\end{equation}
 These are calculated by convolution of the density and the scalar, vectorial or tensorial one-body measures $\omega^{(\nu)}(\bvec{r},\phi)$, which describe the geometry of the hard particles.
 The explicit expression for $\mathcal{F}_\text{ex}[\rho]$ makes use of a truncated and corrected expansion up to rank-two tensors, see Ref.~\cite{annulus} for further details.

In this study we focus on hard discorectangles with rectangular length $L$ and circular diameter $D$ at fixed aspect ratio $p=L/D=10$.
Throughout the manuscript, we consider structures with fixed area fraction $\eta=0.65$, as defined in Eq.~\eqref{eq_eta}.
Typical density profiles in the two geometries are shown in the third row of Fig.~\ref{fig_overview},
which displays the dimensionless total density
\begin{equation}
 \bar{\rho}(\bvec{r}):=\left(LD+\frac{D^2\pi}{4}\right)\!\int_0^{2\pi}\!\frac{\mathrm{d}\phi}{2\pi}\,\rho(\bvec{r},\phi)\,.
 \label{eq_barrho}
\end{equation}
through a color coding and the local orientational director field (representing the locally preferred value of $\phi$) through green arrows.

All structures are calculated by free minimization of the density functional on a spatial grid with resolution $\Delta x=\Delta y=0.2$ and $N_\phi=96$ orientational angles.
Laminar structures are typically initialized by cutting out the inclusions from equilibrium structures in circular confinement.
Then we can also smoothly change the inclusion size ratio $b$ and/or the inclusion distance ratio $c$ to different target values, while continuously minimizing the functional.
To examine the stability of an inclusion tunnel, appropriate structures are superimposed and subsequently minimized for comparison.
To generate comparable structures with different tilt angles for Fig.~\ref{fig_ANGLESDFTbuttonsF},
we also start from two specific structures in circular confinement, possessing eight or nine parallel layers in the central domain.
Then we cut out the two inclusions at typical angles $\alpha$ at which a regular layer structure is maintained and smoothly rotate the inclusions towards other target tilt angles, while continuously minimizing the functional.
Shubnikov structures are initialized either by superimposing a perpendicular domain aligning with the inclusions on equilibrium laminar structures with $\alpha=\pi/2$
or from a random structure with circular orientational director \cite{annulus}.
 After minimization of multiple structures for a given set of parameters, we compare the values of the free energy $\mathcal{F}[\rho]$ to determine the most stable state with minimal free energy.

\subsection{Smectic $\Qvec$-tensor theory \label{app_methods_LdG}}

It is also possible to adapt continuum models to investigate the qualitative behavior of smectics.
Recently, Ref.~\cite{xia-2021-article} proposed a new continuum model, solving for a real-valued smectic order parameter $u$, indicating the local density variation, and a tensor-valued nematic order parameter $\Qvec$.
A detailed discussion about deriving the continuum model can be found in Ref.~\cite{xia-phdthesis}.

Specifically, we use the two-dimensional version of the $\Qvec$-tensor model from \cite{xia-2021-article} with the volumetric free energy:
    \begin{align}\label{eq:model}
        \mathcal{J}_\text{v} (u, \Qvec) = \int_\Omega \bigg(f_\text{s}(u) & + B\left| \mathcal{D}^2 u + q^2 \left(\Qvec+\frac{\Ivec_2}{2}\right) u \right|^2\!\!\!\!\!\!\!\!\cr
    &+f_\text{n}(\Qvec,\nabla \Qvec) \bigg),\!\!\!\!\!\!\!\!
    \end{align}
where
\begin{equation}
    f_\text{s}(u) \coloneqq \frac{a_1}{2} u^2 +\frac{a_2}{3} u^3+\frac{a_3}{4} u^4,
\end{equation}
and
\begin{equation}
        f_\text{n}(\Qvec,\nabla \Qvec)
                        \coloneqq \frac{K}{2}|\nabla \Qvec|^2
                            -l \left(\text{tr}(\Qvec^2)\right) + l \left(\text{tr}(\Qvec^2)\right)^2.
\end{equation}
Here, $K$ is the nematic elastic constant, $l$ represents the nematic bulk parameter, $\Ivec_2$ is the $2\times 2$ identity matrix and $a_1,a_2,a_3, B, q$ are given real parameters.
We fix $a_1 = -5$, $a_2 = 0$, $a_3 = 5$, $B = 10^{-5}$, $q=30$ and $l = 2$, similar to the choice in Ref.~\cite{xia-2021-article}.
In Eq.~\eqref{eq:model}, $\mathcal{D}^2$ denotes the Hessian operator, so that the associated Euler--Lagrange equation for $u$ is a fourth-order partial differential equation.
One can intuitively understand the free energy functional $\mathcal{J}$ as a combination of three contributions: the smectic bulk energy $f_\text{s}$, the coupling effect ($B$-term) between the nematic director and smectic layers and the nematic elastic and bulk energies $f_\text{n}$.

In extreme confinement, we cannot expect the hard rods to perfectly satisfy tangential wall anchoring, as represented by Dirichlet boundary conditions.
Therefore, we weakly impose tangential boundary conditions on both inner boundaries (denoted as $\Gamma_{1}$ and $\Gamma_2$) and outer boundary $\Gamma_\text{outer}$ by means of Rapini--Papoular surface anchoring. To this end, an additional anchoring energy is added to Eq.\ \eqref{eq:model}, leading to the following total energy:
\begin{align}
    \label{eq:unify}
    \!\!\!\!\!\!\mathcal{J}(u, \Qvec) &= \mathcal{J}_\text{v}(u,\Qvec) + \frac{w}{2} \bigg( \int_{\Gamma_\text{outer}} \left| Q-Q_\text{outer}\right|^2\cr
    &\ \ \  + a_\text{r}\big( \int_{\Gamma_{1}} \left| Q-Q_{1} \right|^2
    + \int_{\Gamma_{2}} \left| Q-Q_{2} \right|^2\big)\bigg)\ \
\end{align}
with the prescribed tangential configurations given by
\begin{align}
    &Q_\text{outer} = \begin{bmatrix}
    \frac{y^2}{x^2+y^2}-\frac{1}{2} & -\frac{xy}{x^2+y^2}\\
    -\frac{xy}{x^2+y^2} & \frac{x^2}{x^2+y^2}-\frac{1}{2}
    \end{bmatrix},\\
    &Q_{1} = \begin{bmatrix}
    \frac{y^2}{(x-c/2)^2+y^2}-\frac{1}{2} & -\frac{(x-c/2)y}{(x-c/2)^2+y^2}\\
    -\frac{(x-c/2)y}{(x-c/2)^2+y^2} & \frac{(x-0.3)^2}{(x-c/2)^2+y^2}-\frac{1}{2}
    \end{bmatrix},\\
    &Q_{2} = \begin{bmatrix}
    \frac{y^2}{(x+c/2)^2+y^2}-\frac{1}{2} & -\frac{(x+c/2)y}{(x+c/2)^2+y^2}\\
    -\frac{(x+c/2)y}{(x+c/2)^2+y^2} & \frac{(x+c/2)^2}{(x+c/2)^2+y^2}-\frac{1}{2}
    \end{bmatrix}.
\end{align}
Here, $c$ is the inclusion distance ratio as defined in Fig.~\ref{fig_concept},
$w$ denotes the anchoring weight with larger values representing stronger anchoring
and $a_\text{r}$ accounts for the expected curvature dependence of surface anchoring.
Specifically, the choice of $a_\text{r}$ indicates different anchoring strength $w$ on the outer and $a_\text{r} w$ on the inner boundaries, which can affect the resulting final minimizer with the lowest energy.
Accordingly, we have verified that a slightly weaker anchoring strength, $a_\text{r} <1$, on the inner boundary gives a better consistency with experimental results for the two-holed disk problem.
Therefore, we take $a_\text{r}=0.7$ throughout the manuscript,
except for Fig.~\ref{fig_PDDFTbuttonsLdG}, where the focus lies on illustrating the Laminar-Shubnikov transitions using the same anchoring strength variation on both boundaries for each $w$-continuation step and thus $a_\text{r}=1$ is taken for simplicity.

Due to the nonconvexity of $\mathcal{J}$, there typically exist multiple local minimizers. In our work we employ the deflation technique to discover them \cite{farrell-birkisson-2015-article}; in all figures, we plot the minimizer with lowest energy found for different input parameters $K$ and $w$ (specified accordingly in the manuscript) of the energy functional in Eq.~\eqref{eq:unify}.
More details about the model and associated numerical methods can be found in Refs.~\cite{xia-2021-article, xia-esaim} and \cite[Chapters 8-10]{xia-phdthesis}.
Typical solution profiles in the two geometries are shown in the fourth row of Fig.~\ref{fig_overview}.

\end{document}